\documentclass[a4paper]{article}

\usepackage{INTERSPEECH2020}

\usepackage{subcaption}
\usepackage{bbding}
\usepackage{multirow}
\usepackage{hyperref}

\title{Small-Footprint Keyword Spotting with Multi-Scale Temporal Convolution}

\name{Ximin Li, Xiaodong Wei, Xiaowei Qin}

\address{
  CAS Key Laboratory of Wireless-Optical Communications, \\
  University of Science and Technology of China, Hefei, P. R. China
}
\email{\{lxm123, wxd0301\}@mail.ustc.edu.cn, qinxw@ustc.edu.cn}

\begin{document}

\maketitle

\begin{abstract}
  Keyword Spotting (KWS) plays a vital role in human-computer interaction for smart on-device terminals and service robots. It remains challenging to achieve the trade-off between small footprint and high accuracy for KWS task. 
  In this paper, we explore the application of multi-scale temporal modeling to the small-footprint keyword spotting task. We propose a multi-branch temporal convolution module (MTConv), a CNN block consisting of multiple temporal convolution filters with different kernel sizes, which enriches temporal feature space. Besides, taking advantage of temporal and depthwise convolution, a temporal efficient neural network (TENet) is designed for KWS system\footnote{This work was supported the National Key Research and Development Program of China (2018YFA0701603) and Natural Science Foundation of Anhui Province(2008085MF213)}. Based on the purposed model, we replace standard temporal convolution layers with MTConvs that can be trained for better performance. While at the inference stage, the MTConv can be equivalently converted to the base convolution architecture, so that no extra parameters and computational costs are added compared to the base model. The results on Google Speech Command Dataset show that one of our models trained with MTConv performs the accuracy of 96.8\% with only 100K parameters.
\end{abstract}
\noindent\textbf{Index Terms}: keyword spotting,  convolutional neural network, multi-branch convolutional networks

\section{Introduction}
  \label{sec:introdeuction}

  Keyword spotting (KWS) is a task that aims at detecting pre-trained words in a stream of audio. Wake-up word detection, as an application of keyword spotting, has been increasingly popular in recent years. It is commonly used to initiate an interaction with a voice assistant (e.g., “Hey Siri” \cite{Siri}, “Alexa” \cite{Alexa1, Alexa2}, and “Okay Google” \cite{google}) or distinguish simple common commands (e.g., “yes” or “no”) on mobile devices. Since the fact that such tasks usually run on low-resource devices through continuously listening to specific keywords, it remains challenging to implement highly accurate, low-latency, and small-footprint KWS system.

  With the great success of deep learning in speech recognition, Deep Neural Networks (DNNs) have proven to yield efficient small-footprint solutions for KWS in recent years \cite{helloedge, CNN, DeepResidual, Choi2019, Sinc-Convolutions}. In particular, more advanced architectures, such as Convolutional Neural Networks (CNNs), have been applied to solve KWS problems under limited memory footprint as well as computational resource scenarios, showing excellent accuracy. 
  However, 2D CNN-based methods struggle with capturing the dependency between low and high frequencies with the relatively shallow network. To address the problem, some works \cite{Choi2019, Sinc-Convolutions} utilize 1D temporal convolution to extract high-level frequency features. Despite their success, the ability of aggregating both short-term and long-term temporal informative features has not been considered due to fixed kernel size (i.e., $3 \times 3$ for 2D convolution and $9 \times 1$ for 1D convolution). In fact, it is significant to capture temporal information with different scales, because the characteristics of keywords are usually different on the time scale.

  In this paper, we propose a multi-branch temporal convolution module (MTConv), a CNN structure with different kernel sizes, which can replace the standard temporal convolution layer at the training stage. As a building block of the backbone network, the MTConv enriches the temporal feature space of the model, which handles more temporal information leading to better performance. Specifically, the MTConv can be equivalently converted to the origin temporal convolution layer with an enhanced kernel, which is element-wise added by multi-scale kernels correspondingly, so that we can obtain the same output as the model with MTConvs by the equivalent model with standard temporal convolutions. 
  Moreover, for a better trade-off between small-footprint and accuracy, we propose a temporal efficient neural network (TENet), constructed by inverted bottleneck blocks with $1 \times 1$ convolution, $9 \times 1$ depthwise convolution and $1 \times 1$ convolution. We evaluate the proposed model trained with the MTConv on the Google Speech Command dataset\cite{Dataset}. Our contributions are summarized as follows.

  \begin{itemize}
    \item We propose a temporal efficient neural network based on depthwise temporal convolutions for small-footprint KWS.
    \item We further propose a multi-branch temporal convolution module for obtaining abundant multi-scale temporal features, which can be equivalently converted to the standard temporal convolution at the inference stage.
    \item Our base TENet evaluated on Googles Speech Commands dataset achieves the state-of-the-art accuracy of 96.6\% with only 100K parameters. By replacing the standard temporal convolution with MTConv at training, we further improve the accuracy to 96.8\% with same parameters at the inference stage due to the kernel fusion mechanism.
  \end{itemize}

\section{Method}

\subsection{Data Preprocess}
\label{ssec:dataProcess}

At the beginning of data preprocessing, we apply a band-pass filter of 20Hz/4kHz to reduce noise. After that, forty dimensional Mel-Frequency Cepstrum Coefficient (MFCC) frames are constructed with a 30ms window size and 10ms frame shift.
Finally, we feed the MFCC features as input data into the neural network.
Correspondingly, we denote the MFCC representation as 
$\mathbf{I} \in \mathbb{R}^{T \times F}$,
where $F$ represents the dimension of the MFCC feature, and $T$ denote the number of frames.

\subsection{Temporal Efficient Neural Network}

Following the implementation of temporal convolution\cite{Choi2019}, we first treat the input MFCC features as time series, where the feature dimension is equal to the channels of the input feature map. Therefore, all convolutions in the model are along the temporal dimension, avoiding stacking a large number of layers to capture high-level frequency features.
The input feature map can be represented as $\mathbf{X} \in \mathbb{R}^{T \times 1 \times F}$.

\begin{figure}[t]
  \begin{subfigure}[b]{0.18\textwidth}
    \centering
    \includegraphics[scale=0.5]{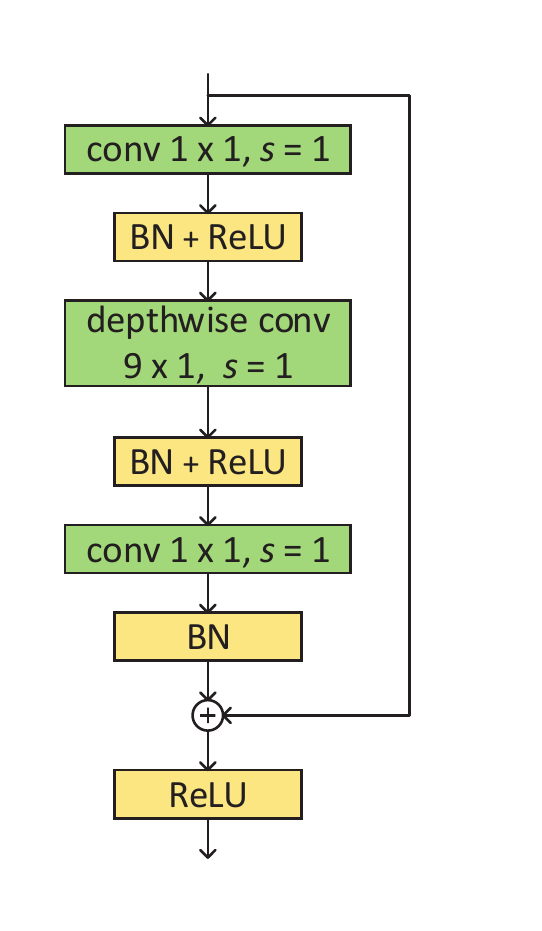}
    \caption{IBB ($stride=1$)}
    \label{fig:IBB_1}
  \end{subfigure}
  \begin{subfigure}[b]{0.3\textwidth}
    \centering
    \includegraphics[scale=0.5]{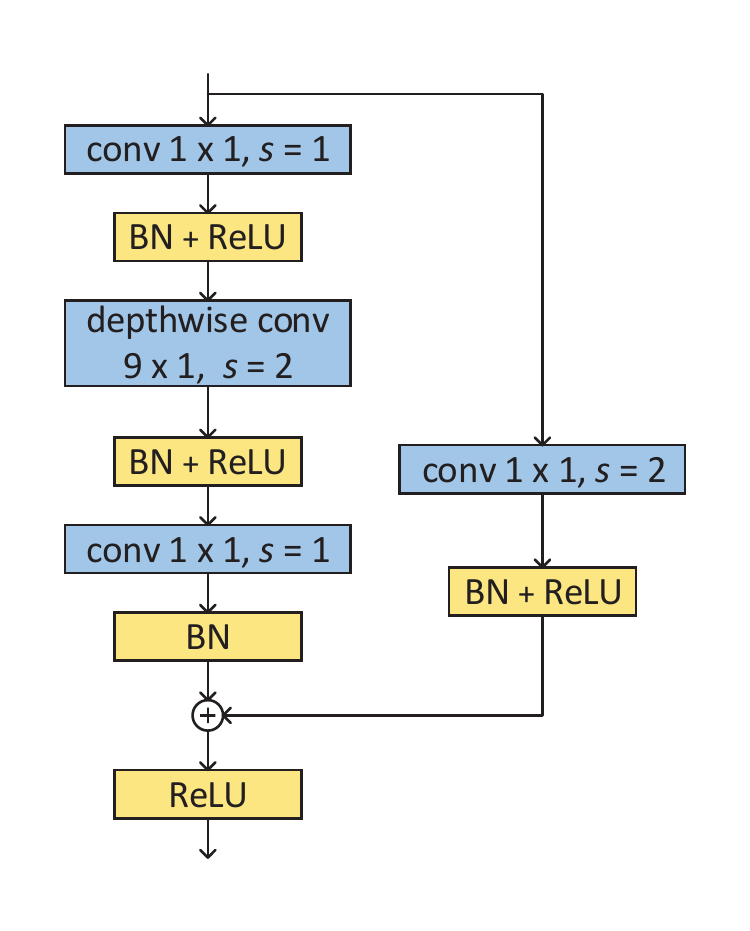}
    \caption{IBB ($stride=2$)}
    \label{fig:IBB_2}
  \end{subfigure}
  \caption{The inverted bottleneck block (IBB) when (a) stride = 1 and (b) stride = 2. BN denotes batch normalization, and `s' indicates stride.}
  \label{fig:IBB}
\end{figure}

Inspired by the concept of inverted residual structure in MobileNetV2\cite{MobileNetV2}, we introduce a bottleneck temporal depth-separable convolution with residuals, denoted as inverted bottleneck block (IBB), which is the key building block of TENet. 
In this paper, the inverted bottleneck block (Figure \ref{fig:IBB}) mainly consists of $1 \times 1$ convolution, $9 \times 1$ depthwise convolution, and $1 \times 1$ convolution. The former $1 \times 1$ convolution aims to embed the input in high-dimensional subspaces by expanding the number of channels, while the latter converts the tensor back to the low-dimensional compact subspaces for information transmission between channels. The $9 \times 1$ depthwise convolution, as a key block of temporal convolution, performs lightweight filtering by applying a single convolutional filter per input channel to accompany a non-linear transformation. The IBB is based on a inverted bottleneck structure where the top and bottom layers are bottleneck layers, and the middle layer is the expansion layer.
We carefully select the ratio between the channel size of the expansion layer and the bottleneck layer as $3$ for a great trade-off between efficiency and performance.
Besides, we use shortcuts\cite{ResNet} directly between the bottleneck layers when the sizes of input and output are matching. Otherwise, we utilize $1 \times 1$ convolution with a stride to match the dimensions. We use ReLU as the nonlinear activation function, and batch normalization\cite{BN} is utilized during training.

\begin{figure}[t]
  \begin{subfigure}{0.2\textwidth}
    \centering
    \includegraphics[scale=0.6]{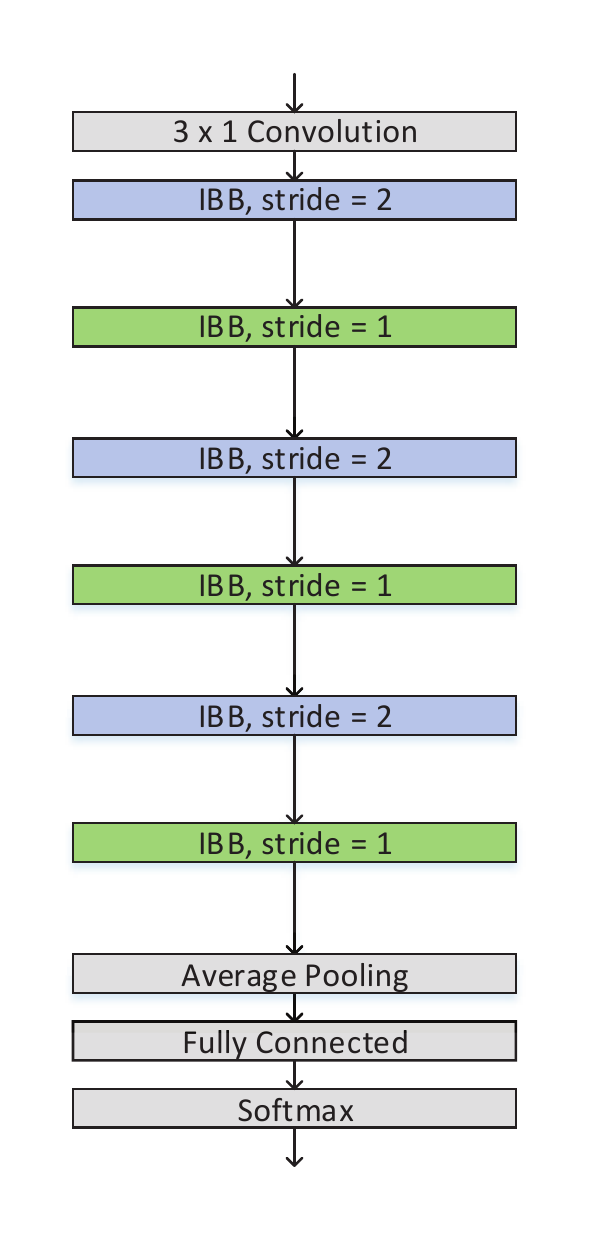}
    \caption{TENet6}
    \label{fig:IBB_1}
  \end{subfigure}
  \begin{subfigure}{0.2\textwidth}
    \centering
    \includegraphics[scale=0.6]{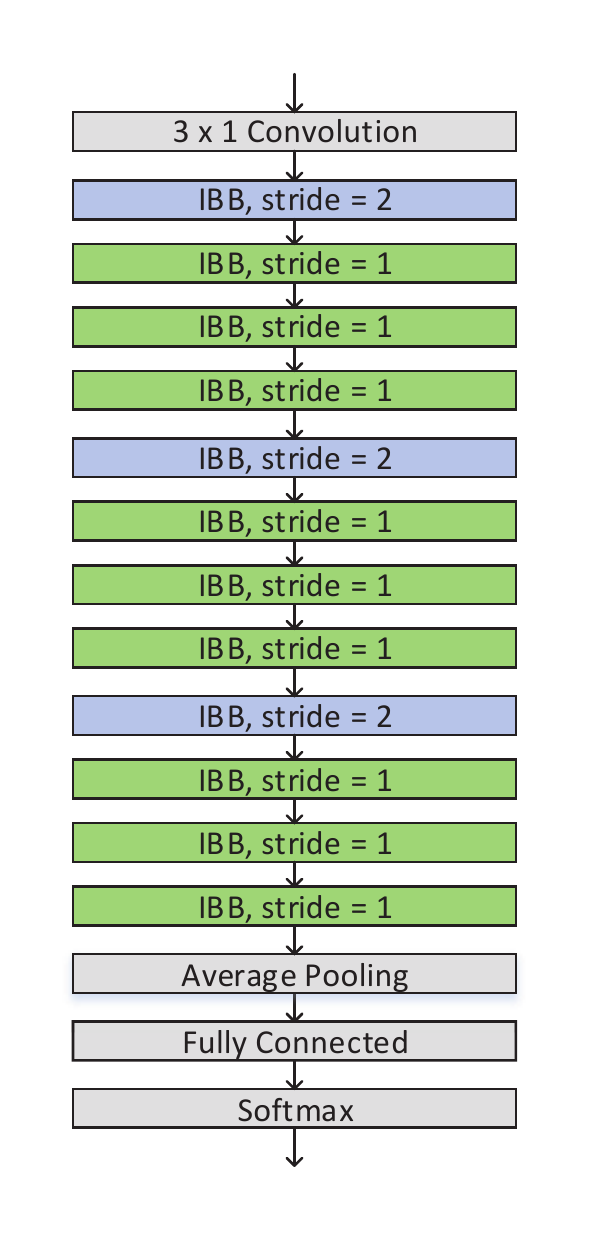}
    \caption{TENet12}
    \label{fig:IBB_2}
  \end{subfigure}
  \caption{two implementations of temporal efficient neural network (TENet) when (a) TENet6 and (b)  TENet12.}
  \label{fig:TENET}
\end{figure}

The temporal efficient neural network can be easily constructed by stacking IBB layers. In Figure \ref{fig:TENET}, we display two TENet implementations, denoted as TENet6 and TENet12, respectively. Both of them start with a $3 \times 1$ convolution with batch normalization and ReLU and end up with the module of the average pooling layer, the full-connected layer, and the softmax layer. Each IBB channel number in TENet6 and TENet12 is 32. For obtaining compact models, we also design TENet6-narrow and TENet12-narrow, whose IBB channel number is 16, and other structural parameters stay constant.

\subsection{Multi-Scale Temporal Convolution}
\subsubsection{Multi-branch Temporal Convolution Module}
\begin{figure*}[t]
  \centering
  \includegraphics[scale=0.58]{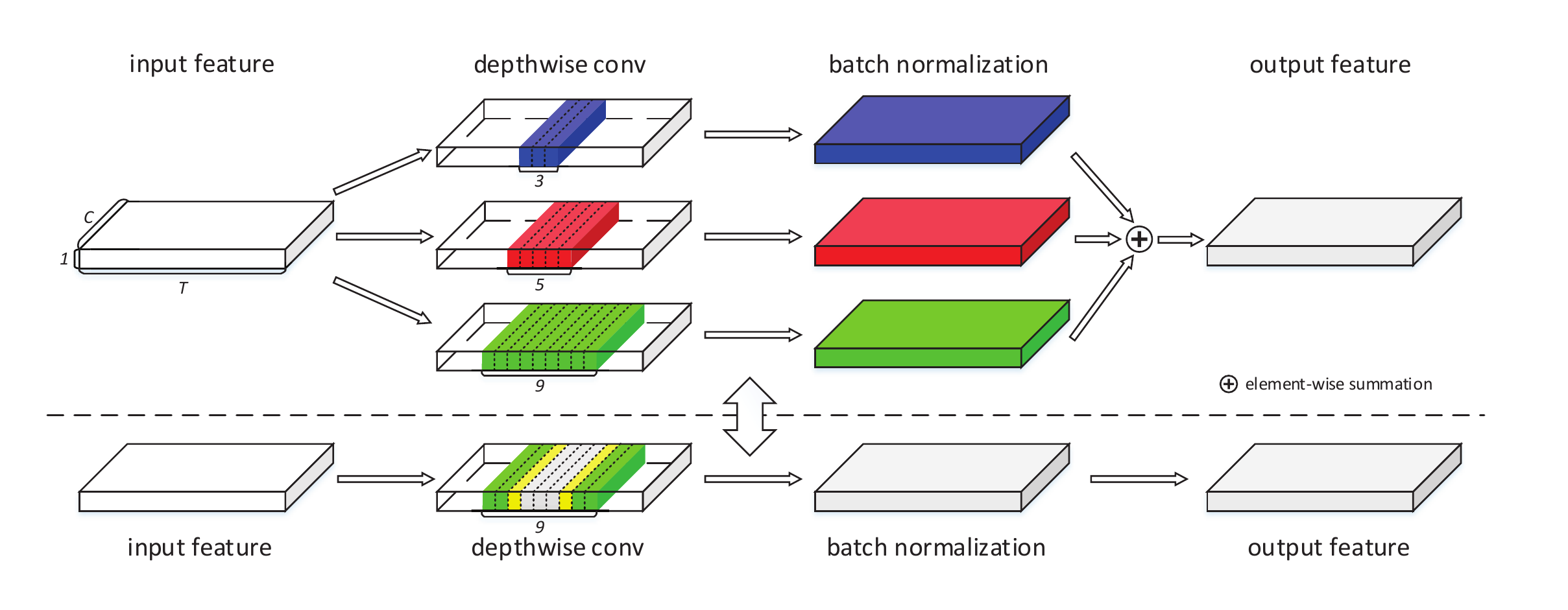}
  \caption{Above the dotted line is a multi-branch temporal convolution module (MTConv) with $3 \times 1$, $5 \times 1$ and $9 \times 1$ kernels, while below the dotted line is a standard depthwise convolution with batch normalization. The two architectures can be converted to each other equivalently.}
  \label{fig:MSEK}
\end{figure*}
To better aggregate both short-term and long-term temporal informative features, we introduce a multi-branch temporal convolution module (MTConv) to strengthen the standard depthwise convolution layer in the proposed model. As a result, compared to the base TENet, we achieve better accuracy by replacing depthwise convolutions in TENet with MTConvs at the training stage. However, at the inference stage, the TENet trained with MTConv can be equivalently converted to the base TENet, which saves model parameters and computational costs.

We implement the MTConv by adopting an architecture of multi-branch depthwise convolutions with different kernel sizes and batch normalization layers to displace the standard depthwise convolution with batch normalization for obtaining richer temporal information.
Concretely, the kernel size of each branch in MTConv differs from each other, thus helping the branch learn unique patterns from its own time granularity.
Figure \ref{fig:MSEK} showing one implementation of MTConv, compared to the standard depthwise convolution with $9 \times 1$ kernel, the MTConv adds two new branches with $5 \times 1$ and $3 \times 1$ kernels, respectively, which enhances the ability of capturing short-term temporal features. 
Considering the feasibility of fusing multi-scale kernels, we element-wise add outputs from the branches to form the ReLU layer’s input instead of concatenating outputs of the branches.

In this work, all the depthwise convolution layers in TENet are replaced with MTConvs while training. The branch number and the sizes of kernels are both selective.
We obtain an enhanced kernel fused by multi-scale kernels from the MTConv, with which we can acquire the same output as the MTConv. The detailed mathematical derivation is in Section \ref{ssec:kernelFusion}. Compared to the kernel trained with the standard convolution, the enhanced kernel contains richer temporal information, resulting in better performance. However, at the inference stage, TENet with MTConv has the same number of parameters and the same computation costs as the base TENet by converting the MTConv to the equivalent standard depthwise convolution.

\subsubsection{Kernel Fusion of MTConv}
\label{ssec:kernelFusion}
For a depthwise convolution with a kernel size of $D \times 1$ and $C$ filters that takes a $C$-channel feature map as input, we denote the concatenation of $C$ filters as $\mathbf{F} \in \mathbb{R}^{D \times 1 \times C}$, the input as $\mathbf{M} \in \mathbb{R}^{T \times 1 \times C}$, and the $C$-channel output as $\mathbf{O} \in \mathbb{R}^{T \times 1 \times C}$ with the convolution of the stride of one and zero padding.
Therefore, for the $j$-th filter at such a layer, the $t$-th element of the corresponding output channel after batch normalization is
\begin{equation}
  \mathbf{O}_{t, 1, j} = \left(
    \sum_{i=-k}^{k}{
      \mathbf{M}_{t+i, 1, j}
      \mathbf{F}_{i+k+1, 1, j}
    } - {\mu}_{j}
  \right) \frac{\gamma_j}{\sigma_j} + \beta_j \,,
  \label{eq:1}
\end{equation}
where we define $k = \frac{D-1}{2}$ upon the assumption that all the kernel sizes are odd.
$\mathbf{M}_{t+i, 1, j}$ is the $(t+i)$-th element in the $j$-th channel of $\mathbf{M}$, and $\mathbf{F}_{i+k+1, 1, j}$ is the $(i+k+1)$-th element of the $j$-th filter. If $t+i < 1$ or $t+i > T$, the corresponding element $\mathbf{M}_{t+i, 1, j}$ is zero. 
$\mu_{j}$ and $\sigma_{j}$ are the values of channel-wise mean and standard deviation of batch normalization, $\gamma_{j}$ and $\beta_{j}$ are the trainable parameters for scaling and shifting, respectively.

Considering a multi-branch temporal convolution module with three branches, we denote the filter, the output and the batch normalization parameters of the i-th branch as 
$\mathbf{F}^{(i)} \in \mathbb{R}^{{D_i} \times 1 \times C}$, $\mathbf{O}^{(i)}$, and$\{\mu^{(i)}, \sigma^{(i)}, \gamma^{(i)}, \beta^{(i)}\}$, respectively, where $D_{i}$ satisfies that $D_{i} = 2k_{i} + 1$.
In order to fuse the three kernels to an enhanced kernel, we do zero padding for small kernels according to Equation (\ref{eq:2}) so that all the kernels are in the same size,
\begin{equation}
  \hat{\mathbf{F}}_{i, 1, j}^{(n)} = 
  \left\{
    \begin{array}{lr}
      \dfrac{\gamma_j}{\sigma_j}
      {\mathbf{F}}_{i - k_m + k^{(n)}, 1, j}^{(n)},  \\
      \quad\, \text{if} \; {k_{m} - k^{(n)} + 1}  \le & i \;\; \le {k_{m} + k^{(n)} + 1}. \\
      0, &\text{otherwise}.\\
    \end{array}
  \right.
  \label{eq:2}
\end{equation}
where $k_m = \max{\{k_1, k_2, k_3\}}$, denoted as the largest size of kernels.
In this case, we can treat the $i$-th branch depthwise convolution with batch normalization as a depthwise convolution with the kernel of $\hat{\mathbf{F}}^{(n)}$ and the bias of $\hat{\beta}^{(i)} = -\dfrac{\mu^{(i)}\gamma^{(i)}}{\sigma^{(i)}}+\beta^{(i)}$.
Therefore, the element-wise addition of the three branches outputs can be easily calculated, which is
\begin{equation} 
  {\mathbf{O}}_{t, 1, j} = 
  \sum_{i=1}^{3}{\mathbf{O}}_{t, 1, j}^{(i)} =
  {\sum_{i=-k_m}^{k_m}}{
    \mathbf{M}_{t+i, 1, j}
    \hat{\mathbf{F}}_{i+k+1, 1, j}
  } + \hat{\beta}_{j} \,,
\end{equation}
where $\hat{\mathbf{F}} = \hat{\mathbf{F}}^{(1)} \oplus \hat{\mathbf{F}}^{(2)} \oplus \hat{\mathbf{F}}^{(3)}$ and
$\hat{\beta} = \hat{\beta}^{(1)} + \hat{\beta}^{(2)} + \hat{\beta}^{(3)}$. $\oplus$ represents the element-wise addition.
In other words, the MTConv is equivalent to a standard depthwise convolution with the kernel of $\hat{\mathbf{F}}$ and the bias of $\hat{\beta}$, and $\hat{\mathbf{F}}$ is the multi-scale enhanced kernel.

\section{Experiment}

\subsection{Experiment Setup}
We evaluate our method using Google’s Speech Commands Dataset \cite{Dataset}, which contains about 65k one-second-long utterance files of 30 different keywords from thousands of people. Following Google’s implementation, we seek to discriminate among 12 classes: “yes,” “no,” “up,” “down,” “left,” “right,” “on,” “off,” “stop,” “go”, unknown, or silence. The dataset is split into 80\% for training set, 10\% for validation set and 10\% for test set according to the SHA-1 hashed name of the audio files. Following Google’s preprocessing procedure, we randomly add background noise, multiplied with a random coefficient sampled from uniform distribution $U(0, 0.1)$, to training samples with a probability of 0.8. We also perform a random $Y$ms time shift to the audio files, where $Y$ is samples from $U(-100, 100)$.

We train our models using TensorFlow\cite{TensorFlow}. We use Adam optimizer\cite{Adam} whose initial learning rate is 0.01 for 30k training iterations, and every 10k iterations the learning rate decays by 0.1. The standard weight decay is set to 0.00004 and the batch size is set to 100.
We train all four implements of TENet with MTConv and the standard convolution, respectively, where the default MTConv is composed of four branches with $3 \times 1$, $5 \times 1$, $7 \times 1$ and $9 \times 1$ convolution kernels.
Each implementation is trained 20 times for an average performance.

Accuracy is the main metric to evaluate models. 
We also take model parameters and multiply operations in the feedforward inference pass into consideration, and plot receiver operating characteristic (ROC) curves of different models.

\subsection{Results and Discussion}

\subsubsection{Comparison with Prior Methods}
In Table \ref{tab:1}, TENets trained without MTConv have been proven to be efficient with the IBB at the comparison with recent methods.
Compared to res15\cite{DeepResidual}, our TENet6-narrow model achieves similar performance with a $14\times$ reduction of parameters and a $1617\times$ reduction of multiply operations. Maintaining a comparable accuracy to TC-ResNet8-1.5 and TC-ResNet14\cite{Choi2019}, TENet12-narrow produces a $4\times$ reduction in parameters and a $2\times$ reduction in multiply operations. For the best performance, TENet12 achieves the same accuracy of 96.6\% as TC-ResNet14-1.5, while the former only requires one-third of parameters and fewer multiply operations. It shows the efficiency of the IBB in temporal convolution.

\begin{table}[t]
  \caption{Comparison of TENets and the baseline models.}
  \label{tab:1}
  \centering
  \begin{tabular}{lccc}
    \toprule
    Model                             &Param.& Mult. &  Acc. \\
    \midrule
    res8-narrow\cite{DeepResidual}    &  20K & 5.65M & 90.1\%\\
    res15-narrow\cite{DeepResidual}   &  43K & 160M  & 94.0\%\\
    res8\cite{DeepResidual}           & 111K & 30M   & 94.1\%\\
    res15\cite{DeepResidual}          & 239K & 894M  & 95.8\%\\
    \midrule
    DS-CNN-S\cite{helloedge}          &  24K & 5.4M  & 94.4\%\\
    DS-CNN-M\cite{helloedge}          & 140K & 19.8M & 94.9\%\\
    DS-CNN-L\cite{helloedge}          & 420K & 56.9M & 95.4\%\\
    \midrule
    TC-ResNet8\cite{Choi2019}         &  66K & 1.12M & 96.1\%\\
    TC-ResNet8-1.5\cite{Choi2019}     & 145K & 2.20M & 96.2\%\\
    TC-ResNet14\cite{Choi2019}        & 137K & 2.02M & 96.2\%\\
    TC-ResNet14-1.5\cite{Choi2019}    & 305K & 4.13M & \textbf{96.6\%}\\
    \midrule
    TENet6-narrow                     &\textbf{17K} & \textbf{553K}  & 96.0\%\\
    TENet12-narrow                    &  31K & 895K  & 96.3\%\\
    TENet6                            &  54K & 1.68M & 96.4\%\\
    TENet12                           & 100K & 2.90M & \textbf{96.6\%}\\
    \bottomrule
  \end{tabular}
\end{table}

\subsubsection{Impact of Multi-branch Modules}
\label{sec:impact}

\begin{table}[t]
  \caption{Comparison between TENets trained with MTConv and the standard convolution.}
  \label{tab:2}
  \centering
  \begin{tabular}{l|c|c|c}
    \toprule
    Model                         &MTConv      &  Acc. & $\Delta$Acc.        \\
    \midrule
    \multirow{2}*{TENet6-narrow}  &$\times$    & 95.97\% &\multirow{2}*{0.15\%}\\
    ~                             &$\checkmark$& 96.12\% &~                   \\
    \midrule
    \multirow{2}*{TENet12-narrow} &$\times$    & 96.34\% &\multirow{2}*{0.18\%}\\
    ~                             &$\checkmark$& 96.52\% &~                   \\
    \midrule
    \multirow{2}*{TENet6}         &$\times$    & 96.42\% &\multirow{2}*{0.32\%}\\
    ~                             &$\checkmark$& 96.74\% &~                   \\
    \midrule
    \multirow{2}*{TENet12}        &$\times$    & 96.63\% &\multirow{2}*{0.21\%}\\
    ~                             &$\checkmark$& 96.84\% &~                   \\
    \bottomrule
  \end{tabular}
\end{table}
We evaluate the performance of TENets trained with or without MTConv, respectively. Note that, whether utilizing MTConv or not, the training hyperparameters maintain constant. As illustrated in Table \ref{tab:2}, compared to training without MTConvs, the accuracy of each TENet implementation with MTConvs increases by 0.15\% to 0.32\%. Moreover, compared to compact networks, such as TENet6-narrow and TENet12-narrow, the wider networks have a larger capacity for abundant temporal informative patterns. It indicates that training with MTConv enables the model handle richer multi-scale temporal information leading to better performance.

\begin{table}[t]
  \caption{Comparison of different kernel scales in MTConv.}
  \label{tab:3}
  \centering
  \begin{tabular}{cccc}
    \toprule
    Model          &Multi-scale kernels        &  Acc. \\
    \midrule
    TENet12        & $\{9\} \times 1$          & 96.63\%\\
    TENet12        & $\{3, 9\} \times 1$       & 96.73\%\\
    TENet12        & $\{3, 5, 9\} \times 1$    & 96.78\%\\
    TENet12        & $\{3, 5, 7, 9\} \times 1$ & 96.84\%\\
    \bottomrule
  \end{tabular}
\end{table}

We also investigate the impact of branch scales on KWS performance. TENet12 is chosen as the base model, and the MTConvs from one to four branches are selected successively. Each module contains a new kernel as compared to the previous one. It is shown in Table \ref{tab:3} that the model without MTConvs achieves an accuracy of 96.63\%, and every additional branch improves performance slightly. The best performance is from the model composed of MTConv with four branches, which has a 0.21\% higher accuracy than the base model.
The results show that increasing kernel scales enriches the comprehensive temporal feature learning.

\begin{figure}[t]
  \centering
  \includegraphics[scale=0.22]{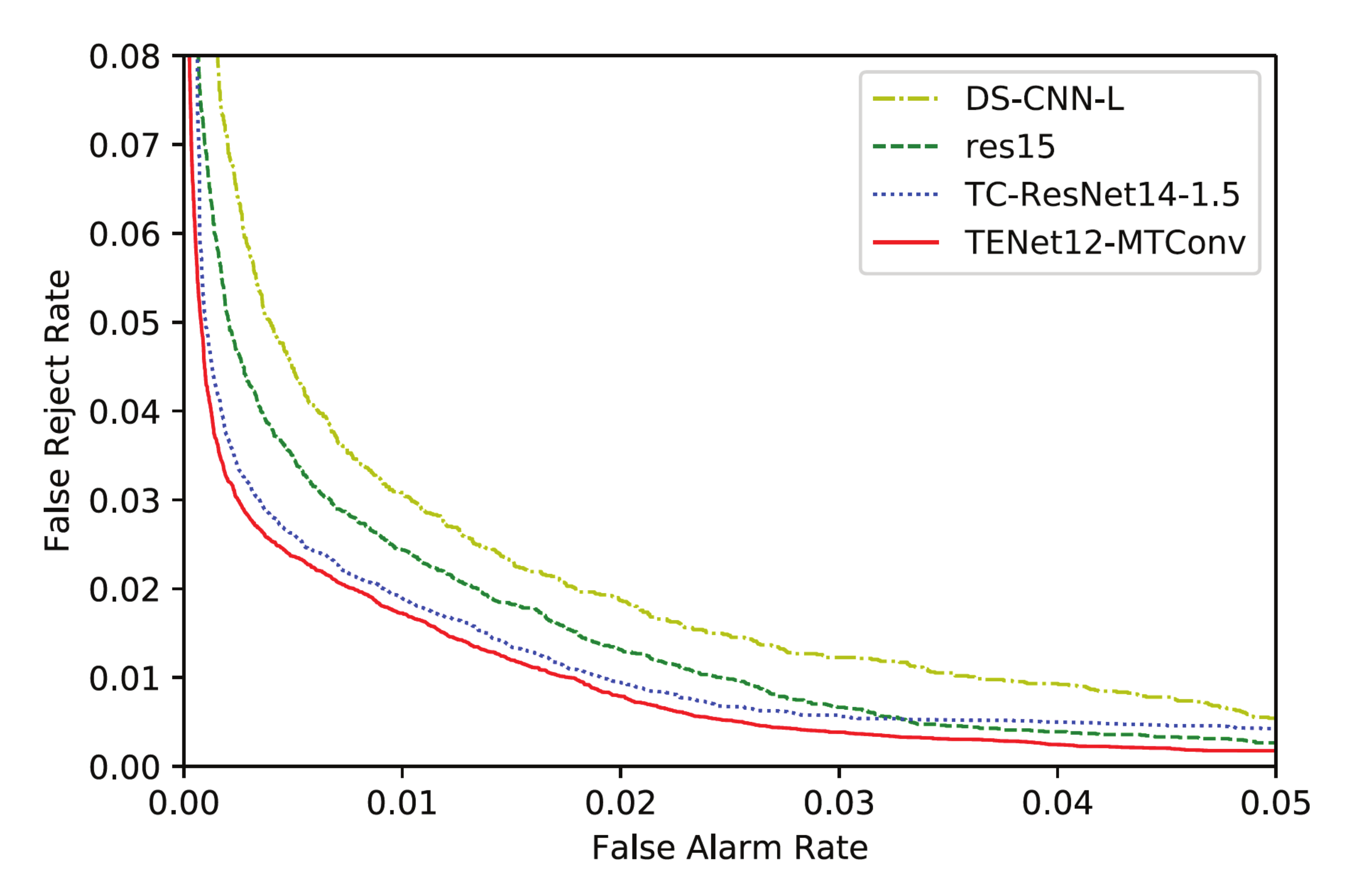}
  \caption{ROC curves for different models}
  \label{fig:roc}
\end{figure}

Moreover, we plot the ROC curves of the best accuracy implements of baseline models as well as the TENet12 trained with MTConv (TENet12-MTConv). As presented in Figure \ref{fig:roc}, TENet12-MTConv outperforms other models at all operating points.

\section{Conclusions}
In this work, we propose a lightweight and efficient model for small-footprint KWS. To simultaneously aggregate short-term and long-term temporal informative features, we introduce a multi-branch temporal convolution module with different kernel sizes, enriching temporal feature space for better performance. Our purposed base model achieves close accuracy of the state-of-the-art model with only one-third of parameters and fewer multiply operations on Google Speech Command Dataset. Additional comparative experiments show the effectiveness of MTConv, with which the purposed model outperforms the base model in performance but no extra parameters or computational costs is added due to the kernel fusion mechanism.

\bibliographystyle{IEEEtran}

\bibliography{template}

\end{document}